\def\be{\begin{equation}}
\def\ee{\end{equation}}
\def\ba{\begin{array}}
\def\ea{\end{array}}

\documentclass[11pt]{article}
\usepackage{amsfonts}
\usepackage{amssymb}
\usepackage{dsfont}
\usepackage{amsmath,amssymb,amstext,amsfonts,array,hhline,tabularx,graphics}
\usepackage{amsthm}

\usepackage{latexsym}
\usepackage{arydshln}
\usepackage{epsfig,graphicx, color}
\usepackage{amsmath}
 \usepackage{cite}
\usepackage{amsthm}
\usepackage{braket}
\graphicspath{{photography/}}{\large }
\usepackage{subcaption}
\usepackage{graphicx}
\usepackage{pgfplots}
\usepackage{inputenx}
\usepackage{epstopdf}

\theoremstyle{plain}
\begin{document}
\parskip=3pt
\parindent=18pt
\baselineskip=20pt \setcounter{page}{1}

 \title{\large\bf Quantum discord of mixed states under noisy channels in the curved spacetime }
\date{}

\author{Yuxuan Xiong, Zhiling Pi, Tinggui Zhang, Xiaofen Huang$^{\ast }$ \\[10pt]
\footnotesize
\small School of Mathematics and Statistics, Hainan Normal University, Haikou 571158, China\\}
\date{}

\maketitle

\centerline{$^\ast$ Correspondence to  huangxf1206@163.com  }
\bigskip

\begin{abstract}
We focus our attention on two-qubit mixed states as initial states, and apply the geometric measure of quantum discord to investigate quantum discord properties in the background of a Schwarzschild black hole under phase damping, phase flip and bit flip channels, respectively. Several analytical complementary relationships based on quantum discords for bipartite subsystems are proposed. For the three channel noises, the behaviors of discords are similar, the accessible discords always degrade as the Hawking acceleration rising, but sudden death never occurs, while the inaccessible discords increase from zero monotonically. Interestingly, in the case of the bit flip channel and phase flip channel, the discords perform symmetrically with the decay probability rising.

\end{abstract}

\section{Introduction}
Quantum discord, with fundamental applications and implications for quantum information processing, is a classification for correlations based on quantum measurements, which plays an important role in quantum information theory\cite{app1, app2, app3, app4, hu2020}. It is defined as the minimum difference between the quantum versions of two classically equivalent expressions of mutual information
under projective measurements \cite{def1,def2}.

For a bipartite composite quantum system $\rho$ on quantum system $\mathcal{H}_A\otimes \mathcal{H}_B$, where $A$ and $B$ are the individual subsystems, the quantum discord is defined as follows,
\begin{equation}
D(\rho)=I(\rho)-\min_{\Pi^A}I[\Pi^A(\rho)],
\end{equation}
here the minimum is over von Neumann measurements (one dimensional orthogonal projectors summing to the identity) $\Pi^A=\{\Pi^A_k\}$ on party $A$, and
$$
\Pi^A(\rho)=\sum_k(\Pi^A_k\otimes I^B)\rho(\Pi^A_k\otimes I^B)
$$
is the resulting state after the measurement. $I(\rho)=S(\rho_{A})+S(\rho_B)-S(\rho)$ is the quantum mutual information,$S(\rho)=-\textrm{Tr}\rho \log_2\rho$ is the Von Neumann entropy.
The intuitive meaning of quantum discord may be interpreted as the minimal loss of correlations (as
measured by the quantum mutual information) due to measurement. This formulation of quantum discord is equivalent to the original definition of quantum discord by Ollivier and Zurek \cite{olli}.

In the last few years, the investigation of relativistic quantum information primarily concentrates on  non-inertial spacetime. Much attention
has been paid to the study of quantum correlations, such as entanglement\cite{entangle1,entangle2, entangle3, entangle4}, coherence\cite{coherence1, coherence2, coherence3}, nonlocality\cite{mi2025, zhang2023}, steering\cite{steering1, steering2, steering3},  shared between inertial and non-inertial observers by discussing how the Unruh or Hawking effect will influence the degree of correlations. Results showed that they are
reduced due to the loss of information caused by Hawking radiation\cite{liu2019, wu2020,torres2019,liu2018, wang2011}. It is obvious that these results not only
help us understand the key of quantum information, but
also play an important role in the study of the information
paradox and entanglement entropy of black holes.

In this paper, the effect of decoherence is investigated for a bipartite system in non-inertial
frames by using phase damping, flip channels and bit flip channels. Two observers, Alice and Bob, sharing a Werner state in non-inertial frames, they are considered to move with a uniform acceleration.  After evaluated the quantum discords analytically, several trade-off relations for discords between subsystems are established, it is explicitly shown that there are some complementarities between the physical accessible and physical inaccessible discords. Aim to describe the dynamics of discords clearly, we plot the discords as a function of acceleration parameter, state parameter and decay parameter. It is clearly to find out both Unruh effect and channel noises can influence the behavior of discords.

The outline of the paper is as follows. In Sec. 2 we consider the bipartite Werner state as the initial state, and derive the analytical expressions of geometry measure of quantum discords for subsystems in the curve spacetime, and then investigate the behaviors of both physical accessible and physical inaccessible discords under the influence of Uhruh effect. In Sec. 3 we discuss the evolution of discords under three environment noises. The last section is devoted to a brief summary.

\section{Geometric measure of quantum discord under the effect of Hawking radiation}

At first, we here give a brief review of the Unruh effect of Dirac fields.
Since similar descriptions have been repeated many times \cite{unrh1, unrh2, unrh3} already, we here
only present the main outlines.
From the Unruh effect of Dirac fields, which, from an inertial perspective, describe
a superposition of the Minkowski monochromatic modes
\begin{equation}\label{R1}
\ket{0}\rightarrow |0\rangle_M=\cos r|0\rangle_{I}|0\rangle_{II}+\sin r|1\rangle_{I}|1\rangle_{II},
\end{equation}
and
\begin{equation}\label{R2}
\ket{1}\rightarrow |1\rangle_M=|1\rangle_{I}|0\rangle_{II},
\end{equation}
where the acceleration parameter $r$ is defined by $\cos r=(e^{-2\pi \omega c/a}+1)^{-1/2}$, $a$ denotes the acceleration of the accelerated observers, and $\omega$ represents the frequency of the Dirac particle, $c$ is the speed of light in vacuum. $\{ |n\rangle_{I(II)}\} $ indicate the Rindler modes in Region $I(II)$.

Next, we fucus on considering the quantum discord of mixed states under Unruh effect of Dirac fields.
Quantum discord as a measure of quantum correlations, initially introduced by Ollivier and Zurek \cite{olli} and by Henderson and Vedral \cite{vedral}, is attracting increasing interest \cite{luo2008, zhu2021, libo2021, hunt2019, aaron2019, hu2025}.
Recently, Daki$\acute{C}$  \textit{et al.} \cite{dis2010} proposed the following geometric measure of quantum discord for bipartite states:
\begin{equation}
D_G(\rho)=\min\limits_{\chi\in \epsilon}\|\rho-\chi\|,
\end{equation}
where $\epsilon$ denotes the set of zero-discord states and the geometric quantity $\|\rho-\chi\|^2=\textrm{Tr}(\rho-\chi)^2$ is the square
of Hilbert-Schmidt norm of Hermitian operators.

Since the Pauli operators $\{\sigma_j\}$, where $j=0, 1, 2, 3$, form an orthogonal basis for qubit system, for any two-qubit state $\rho$,
it can be  decomposed into a linear combination based on the operator bases $\{\sigma_j\}$,  that is the Bloch representation as follows
\begin{equation}
\rho=\frac{1}{4}(I_A\otimes I_B+\sum_{i=1}^3 x_i \sigma_i\otimes I_B+\sum_{j=1}^3 y_jI_A\otimes \sigma_j+\sum_{i,j=1}^3t_{ij}\sigma_i\otimes \sigma_j),
\end{equation}
with $I_A (I_B)$ being the identity matrices of subsystem $A$ ($B$) and  $x_i=\textrm{Tr}\rho (\sigma_i\otimes I_B)$, $y_j=\textrm{Tr}\rho (I_A\otimes \sigma_j)$, $t_{ij}=\textrm{Tr}\rho (\sigma_i\otimes \sigma_j)$ denoting  the correlation coefficients. Therefore,
the geometric measure of quantum discord of a two-qubit state $\rho$ is evaluated as \cite{dis2010}
\begin{equation}\label{discord}
D(\rho)=\frac{1}{4}(\| \textbf{x}\|^2+\|T\|^2-\lambda_{max}),
\end{equation}
where $\textbf{x}:=(x_1, x_2, x_3)^{t}$ is named the Bloch vector, its length is defined by
$\|\textbf{x }\|:=\sqrt{\sum_i x_i^2}$, while $T:=(t_{ij})$ is the correlation  matrix with size $3\times 3$, and $\lambda_{max}$ is the largest eigenvalue of the
matrix $\textbf{x}\textbf{x}^{t}+TT^{t}$. Here the superscript $t$ denotes transpose of
vectors or matrices.

In this article, we primarily employ the geometric measure of quantum discord by introducing accelerating and noise channels to two-qubit mixed states and in both inertial and non-inertial frames.
The analysis reflects the variation of quantum correlation based on geometric measure of quantum discord.
Let the two observers, i.e. Alice and Bob, both the accelerated observers
moving with uniform acceleration, share the following Werner state \cite{werner},
\begin{equation}\label{werner}
\rho_{W}=\frac{1}{6}[(2-p)I_A\otimes I_B+(2p-1)F],
\end{equation}
where state parameter $-1\leq p\leq 1$ and $F$ is the ``flip'' or ``swap'' operator defined by $F({\phi\otimes \psi})=\psi\otimes \phi$. Furthermore,
Werner state (\ref{werner}) has Bloch representation
\begin{equation}
\rho_W=\frac{1}{4}(I_A\otimes I_B+\frac{2p-1}{3}\sum _{i=1}^3 \sigma_i\otimes \sigma_i).
\end{equation}

 In the following, we consider the Werner state for the state parameter $p\in [\frac{1}{2}, 1]$, and assume that both Alice and Bob hover near the event horizon of a Schwarzschild black hole with acceleration $r_a$ and $r_b$, respectively. Due to the Hawking radiation of black hole, the Dirac fields will change from the perspective of the uniformly accelerating observer. So the Werner state $\rho_W$ will be transformed to a four-partite quantum state $\rho_{A_IA_{II}B_IB_{II}}$, whose analytical expression can be obtained using Eqs.(\ref{R1}) and (\ref{R2}), we do not display it here due to its complication.
Since the interior region is causally disconnected from the exterior region of the Schwarzschild black hole, we call the modes inside the event horizon, i.e., the $A_{II}$ and $B_{II}$, the inaccessible modes and the modes outside the event horizon ($A_{I}$ and $B_{I}$) the accessible modes. Since Alice and Bob  cannot access to the Rindler region $II$, we should take trace over modes $ A_{\textit{II}}$ and $ B_{\textit{II}}$, and obtain the partial trace bipartite mixed states $\rho_{A_IB_I}$ with (detail calculation are showed in Appendix ),
\begin{equation}
\begin{aligned}
\rho_{A_IB_I}=&\frac{1}{4}[I_A\otimes I_B-\sin^2 r_a  \sigma_3\otimes I_B-\sin^2 r_b I_A\otimes \sigma_3+\frac{2p-1}{3}\cos r_a\cos r_b\sigma_1\otimes \sigma_1\\
&+\frac{2p-1}{3}\cos r_a\cos r_b\sigma_2\otimes \sigma_2
+(\sin^2 r_a\sin^2 r_b+\frac{2p-1}{3}\cos^2 r_a\cos^2 r_b)\sigma_3\otimes \sigma_3].
\end{aligned}
\end{equation}
Employing  Eq. (\ref{discord}),  we obtain the discord of reduce quantum state $\rho _{A_IB_I}$ as follows,
\begin{equation}
D(\rho _{A_IB_I} )=\frac{(2p-1)^2}{18}\cos^2r_{a}\cos^2r_{b}.
\end{equation}

In the similarly way, we calculate the another three reduce density matrices and the corresponding discords (see detailed calculations in the Appendix), their analytical expressions can be written out explicitly in the following,
\begin{eqnarray}
  D(\rho_{A_IB_{II}}) &=& \frac{(2p-1)^2}{18}\cos^2r_{a}\sin^2r_{b}, \\
  D(\rho_{A_{II}B_I}) &=& \frac{(2p-1)^2}{18}\sin^2r_{a}\cos^2r_{b}, \\
  D(\rho_{A_{II}B_{II}} ) &=& \frac{(2p-1)^2}{18}\sin^2r_{a}\sin^2r_{b}.
\end{eqnarray}

It can be easily checked that there is a trade-off relation among geometric measure of quantum discords $D(\rho_{A_{I}B_I})$ , $D(\rho_{A_{II}B_{I}} )$, $D(\rho_{A_{I}B_{II}} )$ and $D(\rho_{A_{II}B_{II}} )$, that is,
\begin{equation}
D(\rho _{A_IB_I} )+D(\rho_{A_IB_{II}})+D(\rho_{A_{II}B_I})+D(\rho_{A_{II}B_{II}})=\frac{(2p-1)^2}{18}.
\end{equation}
This complementary relationship shows that the total amount of discords from physically accessible and physically inaccessible is a constant.

Furthermore, in order to illustrate them, we plot the geometric measure of quantum discords $D(\rho_{A_IB_{I}})$ and $D(\rho_{A_{II}B_{II}})$ with parameters $r$.
These charts show an interesting phenomenon.
With the increasing in Hawking acceleration $r$, the physically accessible quantum discord monotonically  decrease, however, the physically inaccessible quantum discord is increasing from zero. This interesting phenomenon indicates that the acceleration
noise introduced by Hawking radiation destroys the physically accessible discord between Alice and Bob. This further implies Hawking radiation is equivalent to a kind of heat effect which reduces the physically accessible coherence and of course may breed the physically inaccessible discord.

\begin{figure}
	\centering
	\begin{subfigure}[h]{0.3\textwidth}
		\includegraphics[width=\textwidth]{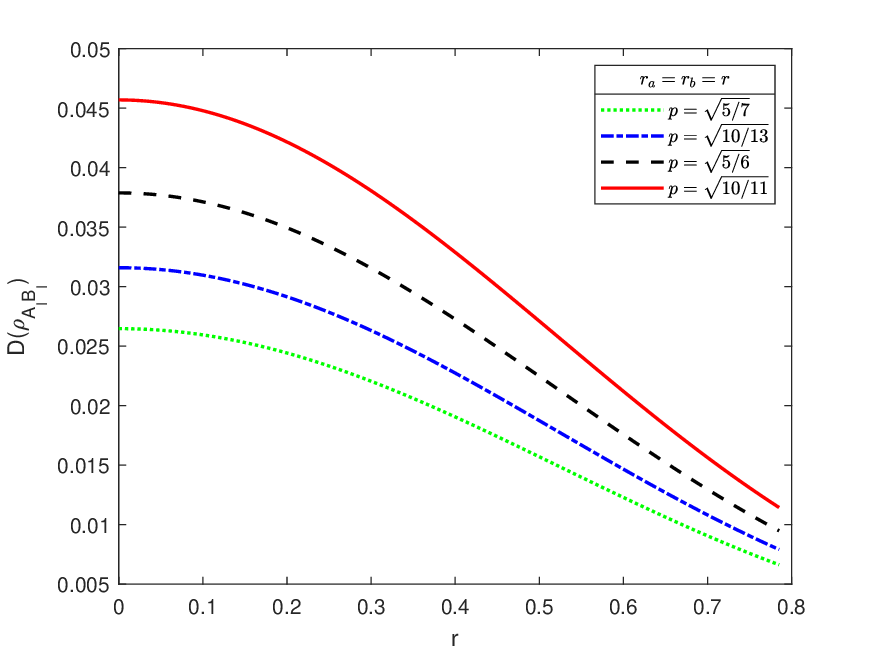}
		\caption{  }
		\label{fig:subfig1}
	\end{subfigure}
	\begin{subfigure}[h]{0.3\textwidth}
		\includegraphics[width=\textwidth]{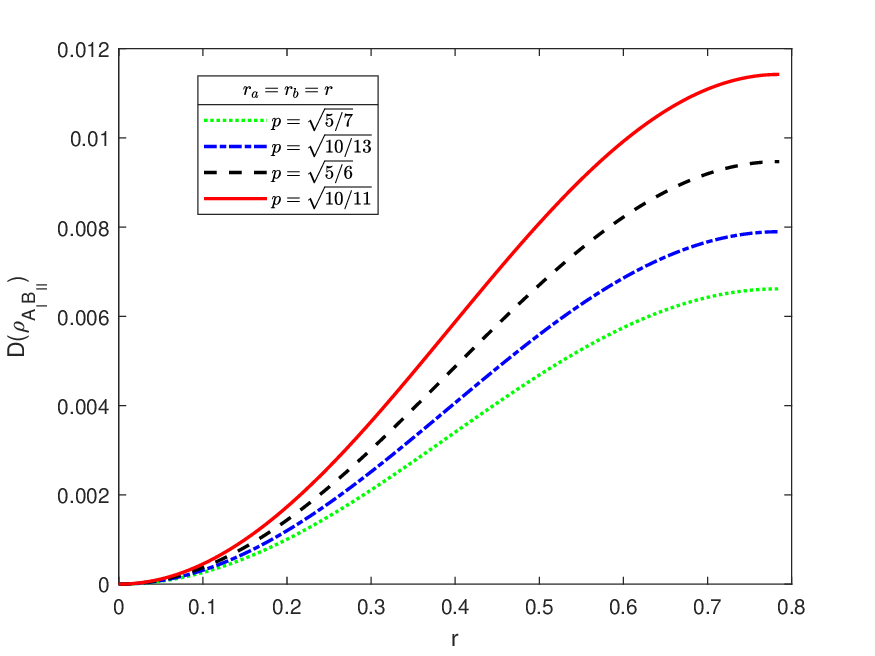}
		\caption{  }
		\label{fig:subfig1}
	\end{subfigure}
  \begin{subfigure}[h]{0.3\textwidth}
		\includegraphics[width=\textwidth]{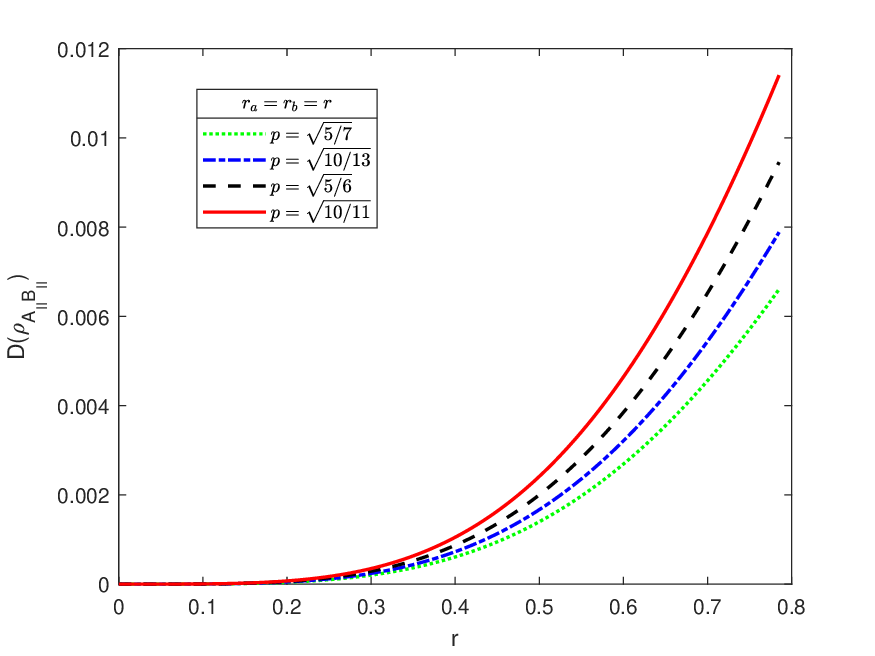}
		\caption{  }
		\label{fig:subfig2}
	\end{subfigure}
	\label{Fig.2}\caption{Plot quantum discords $D(\rho_{A_{I}B_I})$ , $D(\rho_{A_{I}B_{II}} )$ and $D(\rho_{A_{II}B_{II}} )$ when Hawking acceleration $r_a=r_b=r$ for various state parameters.}
\end{figure}

\section{Geometric measure of quantum discord affected by noisy environment  }
We further assume that each subsystem is influenced by its own noisy environment. The interaction between the system and its environment introduces decoherence to the system, which
is a process of undesired correlation between the system and the environment. The evolution of the density matrix of a system in a noisy environment is described in terms of the Kraus
operator formalism. The action of a noisy environment can be described as
\begin{equation}
	\rho \rightarrow \rho^{evo}=\sum_i E_i \rho E_i^{\dagger},
\end{equation}
where $\rho(\rho^{evo})$ is the density matrix of a initial (final) state, $E_i$ ($E_i^{\dagger}$) is the single-qubit Kraus (complex conjugate) operator of the noisy channel. Except the Unruh effect,  let's move to discuss the interconnections among the formulas under various noisy environments, here we focus on phase damping channel, phase flip channel and bit flip channel as examples.

\textbf{In case of phase damping channel}: First of all, we will consider the case of the phase damping channel. The single qubit Kraus operators for the phase
damping channel are given by
\begin{equation}
\begin{aligned}
&E_0=\left(\begin{array}{cc}
		1 & 0 \\
		0 & \sqrt{1-k}
	\end{array}\right),
&E_1=\left(\begin{array}{cc}
		0 & 0 \\
		0 & \sqrt{k}
	\end{array}\right),\\
\end{aligned}
\end{equation}
where $k \in [0, 1]$ is a decay probability and in our study we assume that it depends only on time \cite{decay}.

In order to simplify the calculations, it is assumed that decay probability parameters
corresponding to local coupling of the channel with the
qubits are the same. The geometric measure of quantum discords can be calculated easily by applying Eq. (\ref{discord}) and are given by
\begin{eqnarray}
  D(\rho _{A_IB_I} ) &=& \frac{[(1-k)(2p-1)]^2}{18}\cos^2r_{a}\cos^2r_{b}, \\
  D(\rho_{A_IB_{II}}) &=& \frac{[(1-k)(2p-1)]^2}{18}\cos^2r_{a}\sin^2r_{b}, \\
  D(\rho_{A_{II}B_I}) &=& \frac{[(1-k)(2p-1)]^2}{18}\sin^2r_{a}\cos^2r_{b}, \\
  D(\rho_{A_{II}B_{II}}) &=& \frac{[(1-k)(2p-1)]^2}{18}\sin^2r_{a}\sin^2r_{b}.
\end{eqnarray}

It is easy checked that the geometric measure of quantum discords for reduced quantum states under the effect of phase damping channel satisfied trade-off relation as follows,
\begin{equation}
D(\rho _{A_IB_I} )+D(\rho_{A_IB_{II}})+ D(\rho_{A_{II}B_I})+D(\rho_{A_{II}B_{II}})=\frac{[(1-k)(2p-1)]^2}{18}.
\end{equation}

To illustrate the variation of the discords with respect to the acceleration parameter $r$ and decay probability parameter $k$, we plot the discords for the bipartite systems in Fig. 2. Notice that the physically accessible discord decreases along with the increasing accelerated observers, while the physically inaccessible discord increases as Hawking temperature rising. It should be recognized that the discord for each system has never been disappeared even in the infinite acceleration, it means that the discord sudden death never occurs. Furthermore, we consider the influence of both Hawking radiation and noisy environment interactive impact simultaneously for the decay probability parameter $k=\frac{1}{3}$. Both the noisy environment
and the acceleration of subsystems influence the discord. In case of $p=\frac{1}{2}$ or $k=1$, discords completely vanish.
\begin{figure}[ht]
	\centering
	\begin{subfigure}[h]{0.3\textwidth}
		\includegraphics[width=\textwidth]{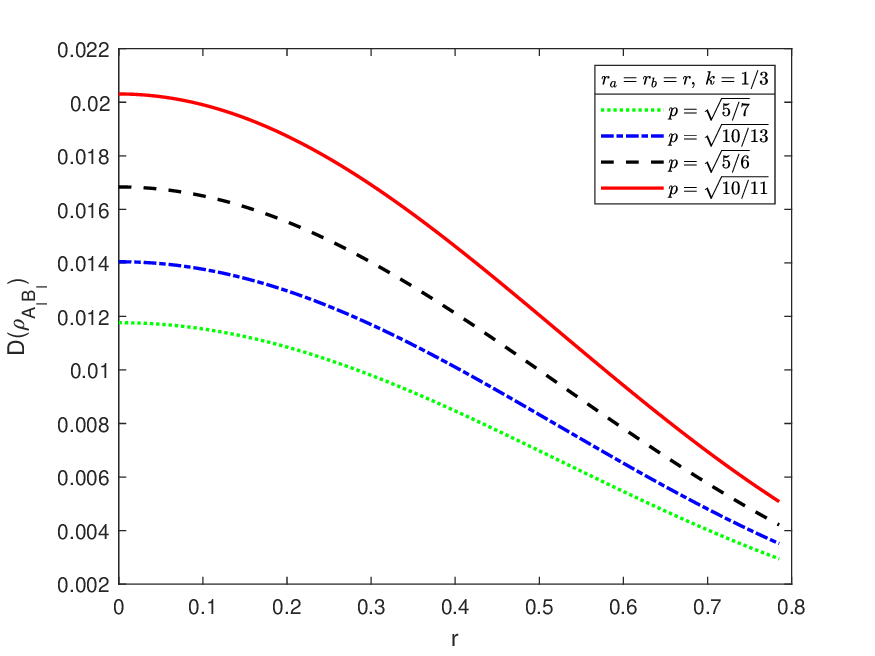}
		\caption{   }
		\label{fig:subfig1}
	\end{subfigure}
  \begin{subfigure}[h]{0.3\textwidth}
		\includegraphics[width=\textwidth]{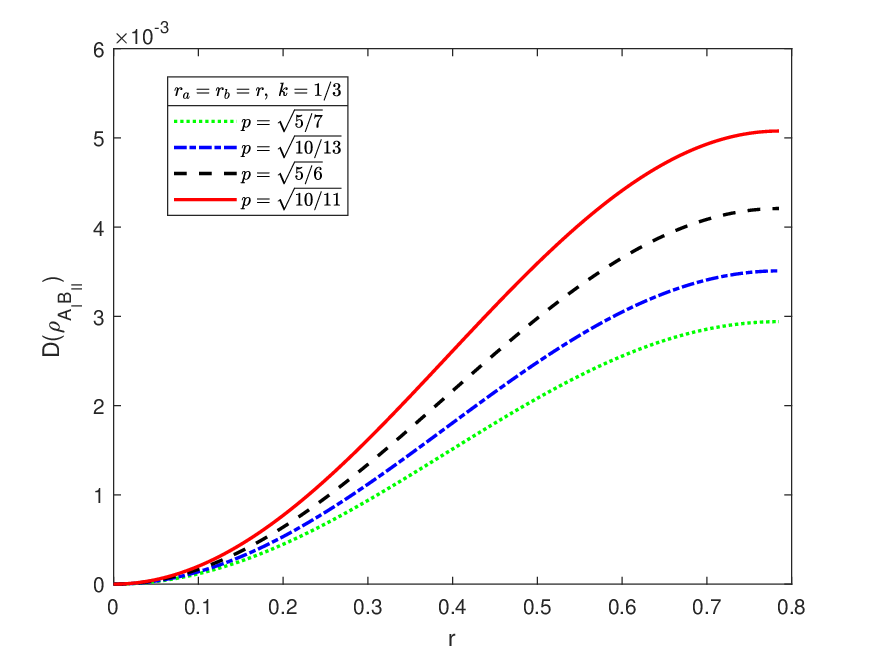}
		\caption{   }
		\label{fig:subfig2}
	\end{subfigure}
  \begin{subfigure}[h]{0.3\textwidth}
		\includegraphics[width=\textwidth]{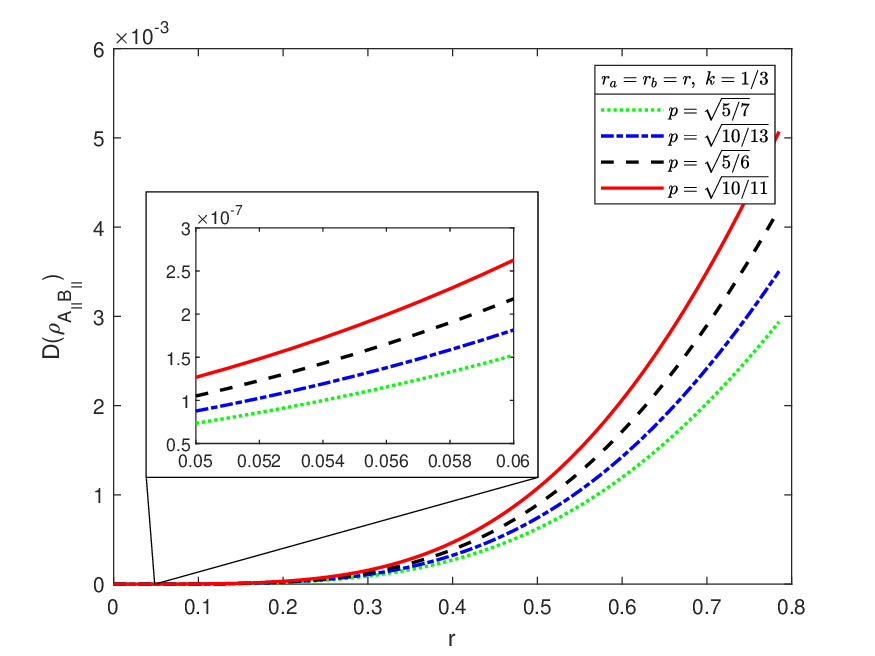}
		\caption{ }
		\label{fig:subfig2}
	\end{subfigure}

	\begin{subfigure}[h]{0.3\textwidth}
		\includegraphics[width=\textwidth]{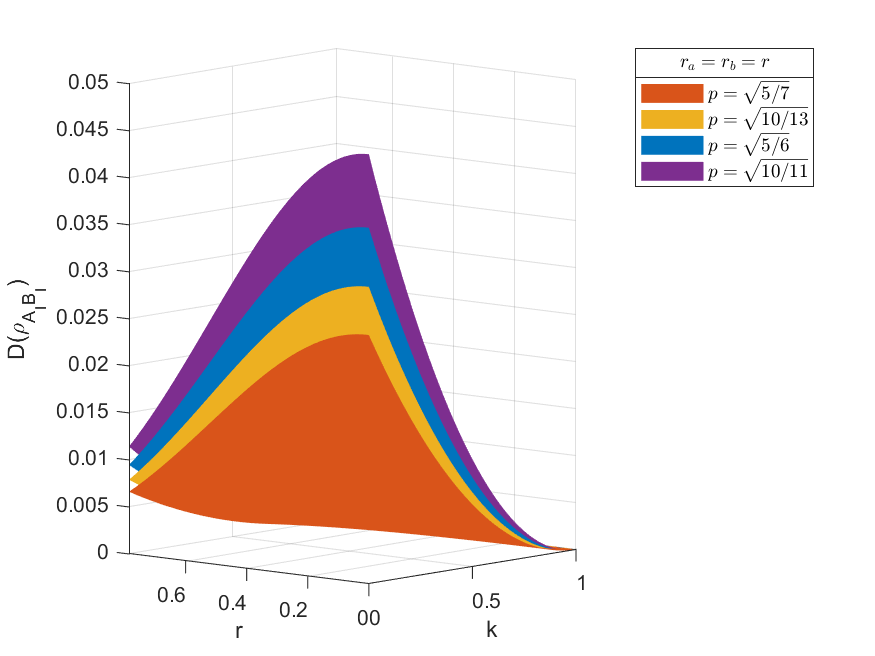}
		\caption{  }
		\label{fig:subfig1}
	\end{subfigure}
  \begin{subfigure}[h]{0.3\textwidth}
		\includegraphics[width=\textwidth]{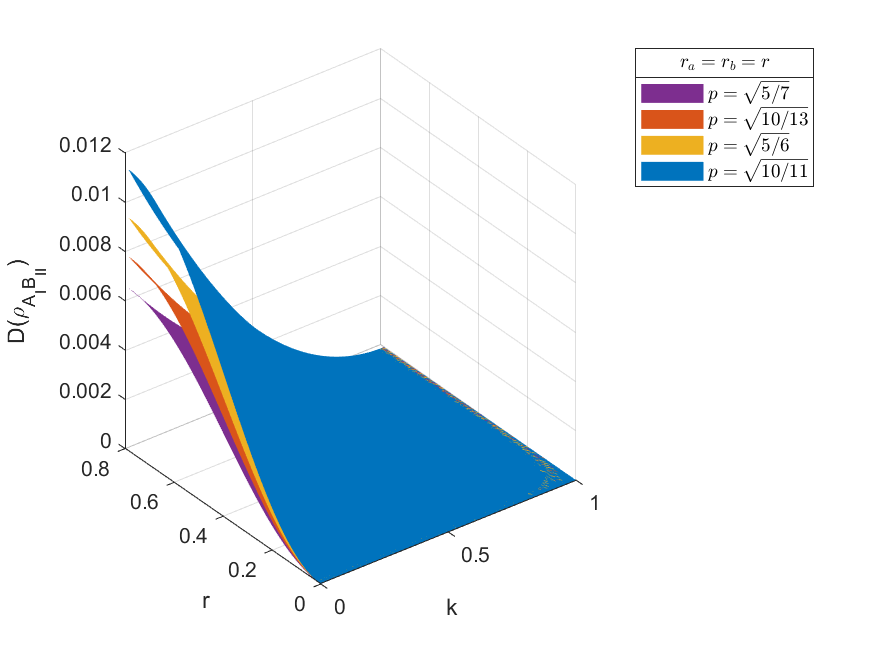}
		\caption{   }
		\label{fig:subfig2}
	\end{subfigure}
  \begin{subfigure}[h]{0.3\textwidth}
		\includegraphics[width=\textwidth]{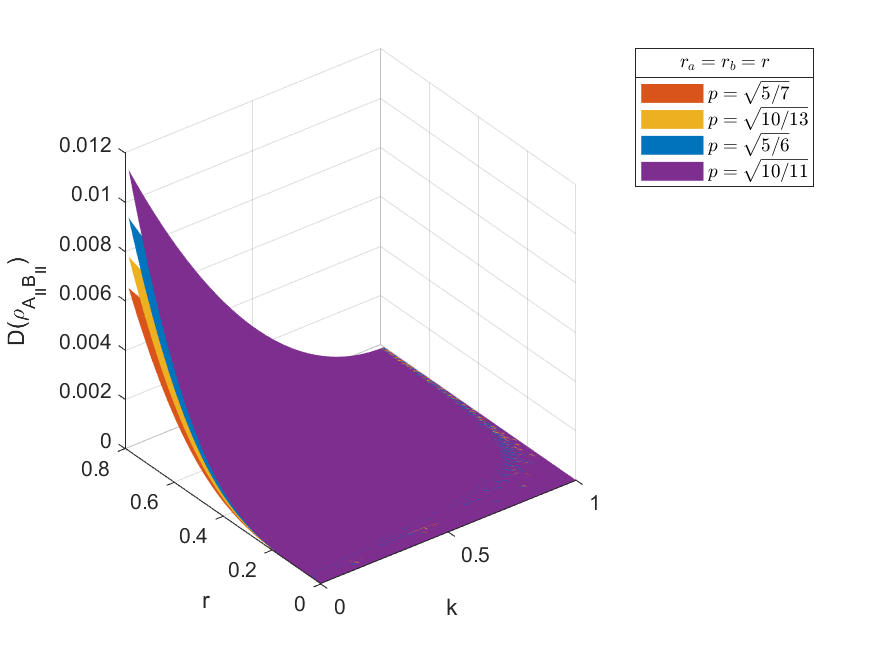}
		\caption{  }
		\label{fig:subfig2}
	\end{subfigure}
	\label{Fig.2}\caption{Plot quantum discords $D(\rho_{A_{I}B_I})$, $D(\rho_{A_{I}B_{II}} )$ and $D(\rho_{A_{II}B_{II}} )$ under phase damping noisy when $r_a=r_b=r$. The upper three subfigures are the cases that discord is a function of acceleration parameter $r$ with several different values of state parameter $p$ for $k=\frac{1}{3}$. The lower three subfigures are the cases that discord is a function of both acceleration parameter $r$ and decay probability parameter $k$.}
\end{figure}

\textbf{In case of phase flip channel}:
When we consider the case of the  phase flip channel, the single qubit Kraus operators are given by,
\begin{equation}
\begin{aligned}
&E_0=\left(\begin{array}{cc}
		\sqrt{1-k} & 0 \\
		0 & \sqrt{1-k}
	\end{array}\right),
&E_1=\left(\begin{array}{cc}
		\sqrt{k} & 0 \\
		0 & -\sqrt{k}
	\end{array}\right).\\
\end{aligned}
\end{equation}

The geometric measure of quantum discords, influenced by the phase
flip noise can be calculated by using the definition as given in Eq. (\ref{discord}), are given by
\begin{eqnarray}
  D(\rho _{A_IB_I}) &=& \frac{(2p-1)^2\big(1-2k)^4}{18}\cos^2r_{a}\cos^2r_{b}, \\
  D(\rho_{A_IB_{II}}) &=& \frac{(2p-1)^2\big(1-2k)^4}{18}\cos^2r_{a}\sin^2r_{b}, \\
  D(\rho_{A_{II}B_I}) &=& \frac{(2p-1)^2\big(1-2k)^4}{18}\sin^2r_{a}\cos^2r_{b}, \\
  D(\rho_{A_{II}B_{II}}) &=& \frac{(2p-1)^2\big(1-2k)^4}{18}\sin^2r_{a}\sin^2r_{b}.
\end{eqnarray}
Furthermore, we derive a trade-off relation about geometric measure of quantum discords for reduced quantum states under the effect of phase flip channel as follows,
\begin{equation}
D(\rho _{A_IB_I})+D(\rho_{A_IB_{II}})+D(\rho_{A_{II}B_I})+D(\rho_{A_{II}B_{II}})=\frac{(2p-1)^2\big(1-2k)^4}{18}.
\end{equation}

In Fig. 3, we show geometric measure of quantum discord as a function of
the acceleration parameter $r$, decay probability parameter $k$ and state parameter $p$. We plot quantum discords with  $r=r_a=r_b$ and $k=\frac{1}{3}$ and for the phase damping channel. The geometric measure of quantum discord decreases as the decoherence parameter $k$ $(0<k<1)$ increases and vanishes completely when $k = \frac{1}{2}$.  Discords completely vanish when $p=\frac{1}{2}$.
\begin{figure}[h]
	\centering
\begin{subfigure}[h]{0.3\textwidth}
		\includegraphics[width=\textwidth]{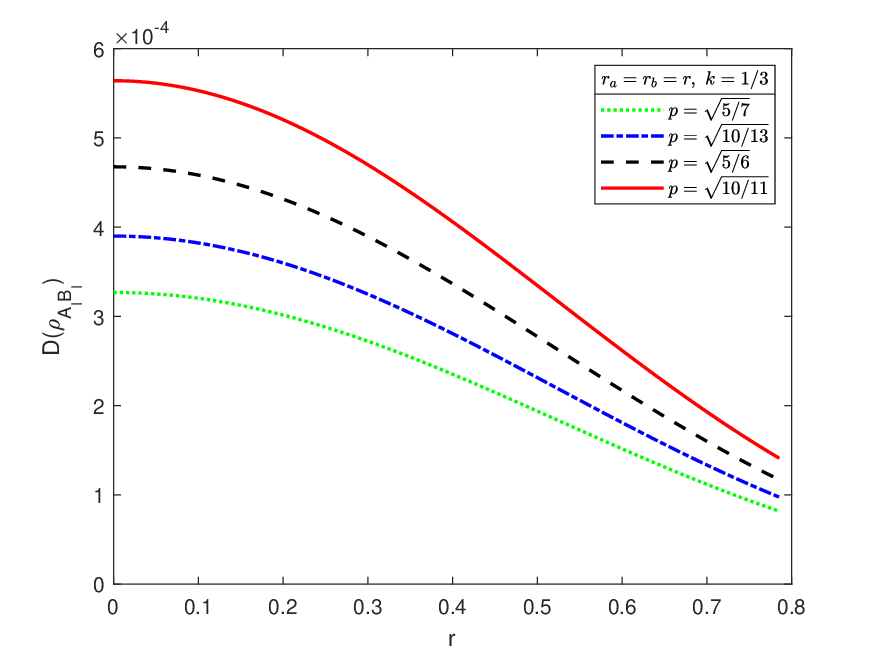}
		\caption{ }
		\label{fig:subfig1}
	\end{subfigure}
  \begin{subfigure}[h]{0.3\textwidth}
		\includegraphics[width=\textwidth]{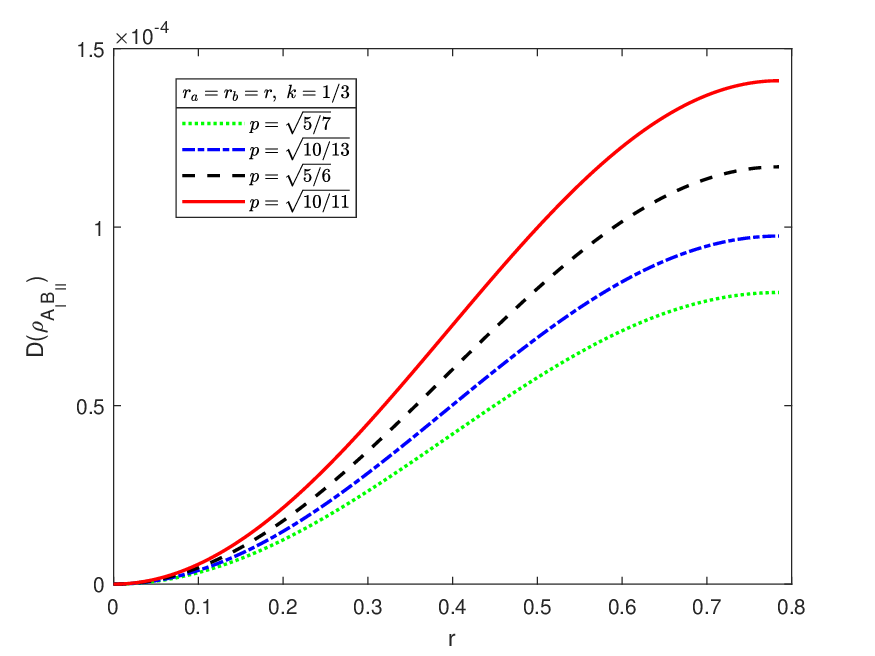}
		\caption{  }
		\label{fig:subfig2}
	\end{subfigure}
  \begin{subfigure}[h]{0.3\textwidth}
		\includegraphics[width=\textwidth]{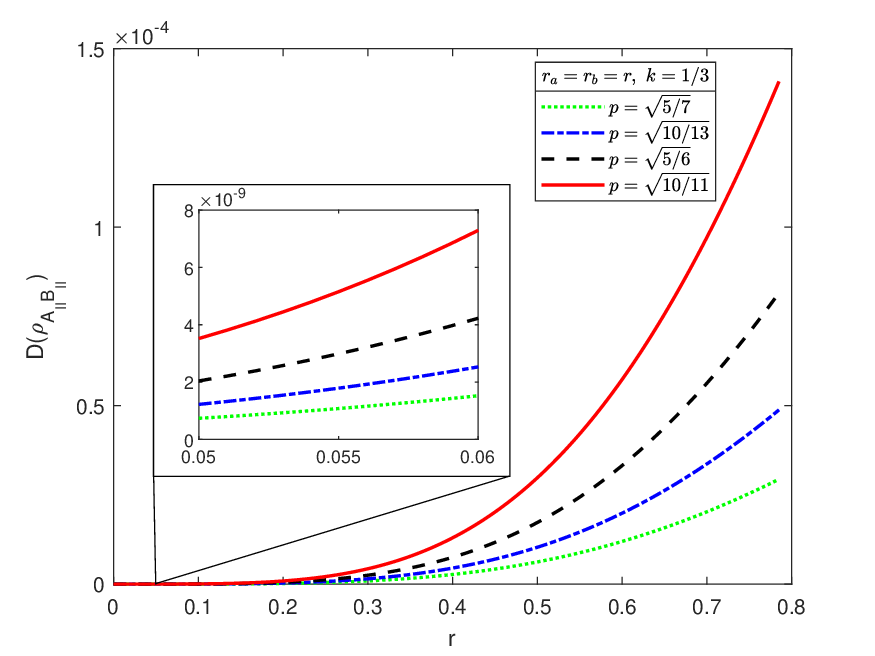}
		\caption{  }
		\label{fig:subfig2}
	\end{subfigure}

\begin{subfigure}[h]{0.3\textwidth}
		\includegraphics[width=\textwidth]{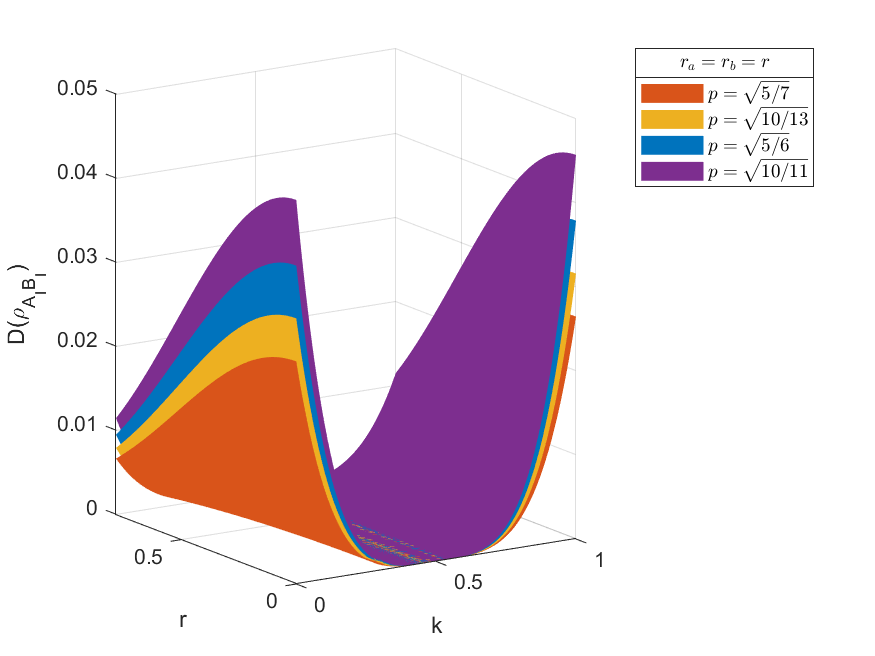}
		\caption{ }
		\label{fig:subfig1}
	\end{subfigure}
  \begin{subfigure}[h]{0.3\textwidth}
		\includegraphics[width=\textwidth]{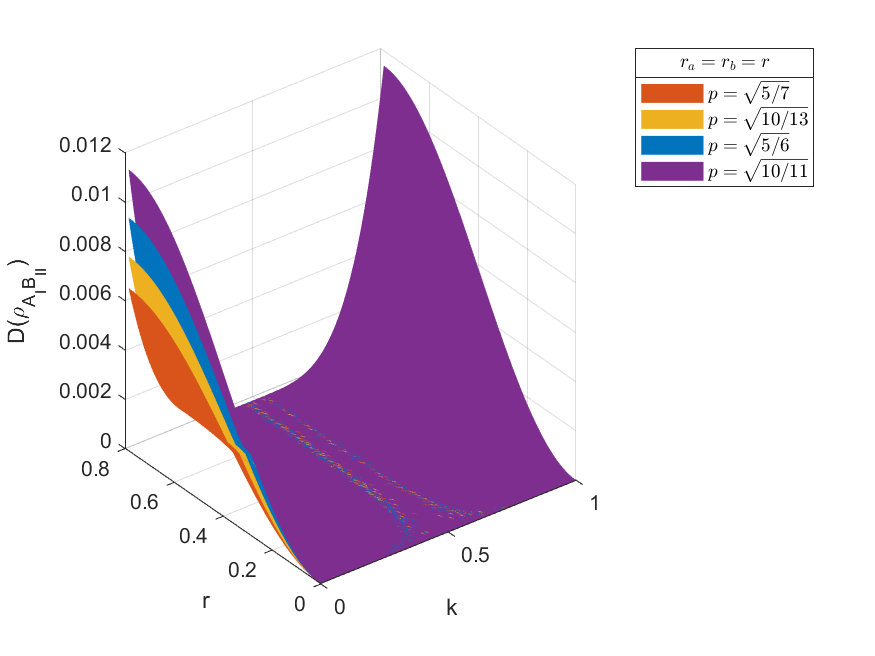}
		\caption{  }
		\label{fig:subfig2}
	\end{subfigure}
  \begin{subfigure}[h]{0.3\textwidth}
		\includegraphics[width=\textwidth]{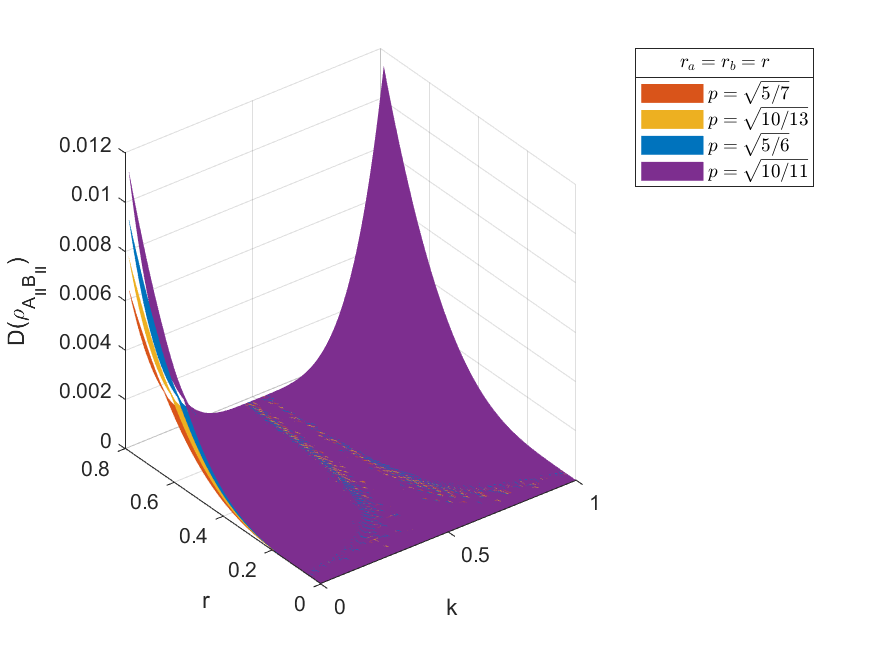}
		\caption{  }
		\label{fig:subfig2}
	\end{subfigure}

	\label{Fig.4}\caption{Plot quantum discords $D(\rho_{A_{I}B_I})$, $D(\rho_{A_{I}B_{II}} )$ and $D(\rho_{A_{II}B_{II}} )$ for the phase flip channel when $r_a=r_b=r$. The upper subfigures are the cases that discord is a function of acceleration parameter $r$  for $k=\frac{1}{3}$. The lower subfigures are the cases that discord is a function of both acceleration parameter $r$ and decay probability parameter $k$. }
\end{figure}

\textbf{In case of bit flip channel}:
We now consider the case of bit flip channel. The single-qubit Kraus operators for
the bit flip channel are given by
\begin{equation}
\begin{aligned}
&E_0=\left(\begin{array}{cc}
		\sqrt{1-k} & 0 \\
		0 & \sqrt{1-k}
	\end{array}\right),
&E_1=\left(\begin{array}{cc}
		0 & \sqrt{k} \\
		\sqrt{k} & 0
	\end{array}\right).\\
\end{aligned}
\end{equation}

Similarly, we can derive the geometric measure of quantum discords influenced by the bit flip channel by using Eq. (\ref{discord}),
\begin{eqnarray}
  D(\rho_{A_IB_I} ) &=& \frac{(2p-1)^2\big(1+(1-2k)^4\big)}{36}\cos^2r_{a}\cos^2r_{b}, \\
  D(\rho_{A_IB_{II}}) &=& \frac{(2p-1)^2\big(1+(1-2k)^4\big)}{36}\cos^2r_{a}\sin^2r_{b}, \\
  D(\rho_{A_{II}B_I}) &=& \frac{(2p-1)^2\big(1+(1-2k)^4\big)}{36}\sin^2r_{a}\cos^2r_{b}, \\
  D(\rho_{A_{II}B_{II}}) &=& \frac{(2p-1)^2\big(1+(1-2k)^4\big)}{36}\sin^2r_{a}\sin^2r_{b}.
\end{eqnarray}
Also, we obtain a trade-off relation about geometric measure of quantum discords for subsystems under the effect of bit flip channel as follows,
\begin{equation}
 D(\rho_{A_IB_I} )+ D(\rho_{A_IB_{II}})+D(\rho_{A_{II}B_I})+D(\rho_{A_{II}B_{II}})=\frac{(2p-1)^2\big(1+(1-2k)^4\big)}{36}.
\end{equation}

In Fig. 4, we show geometric measure of quantum discord as a function of
the acceleration parameter $r$, decay probability parameter $k$ and state parameter $p$. Furthermore, we plot quantum discords in the case $r=r_a=r_b$ and $k=\frac{1}{3}$ for the bit flip channel. The geometric measure of quantum discords increase as the state parameter $p$ $(\frac{1}{2} <p<1)$ increases and vanishes completely when $p = \frac{1}{2}$, interesting, quantum discords exhibit symmetric variation with respect to the decay probability parameter $k$, the discords are zero in case of $k=\frac{1}{2}$.
\begin{figure}[h]
	\centering
\begin{subfigure}[h]{0.3\textwidth}
		\includegraphics[width=\textwidth]{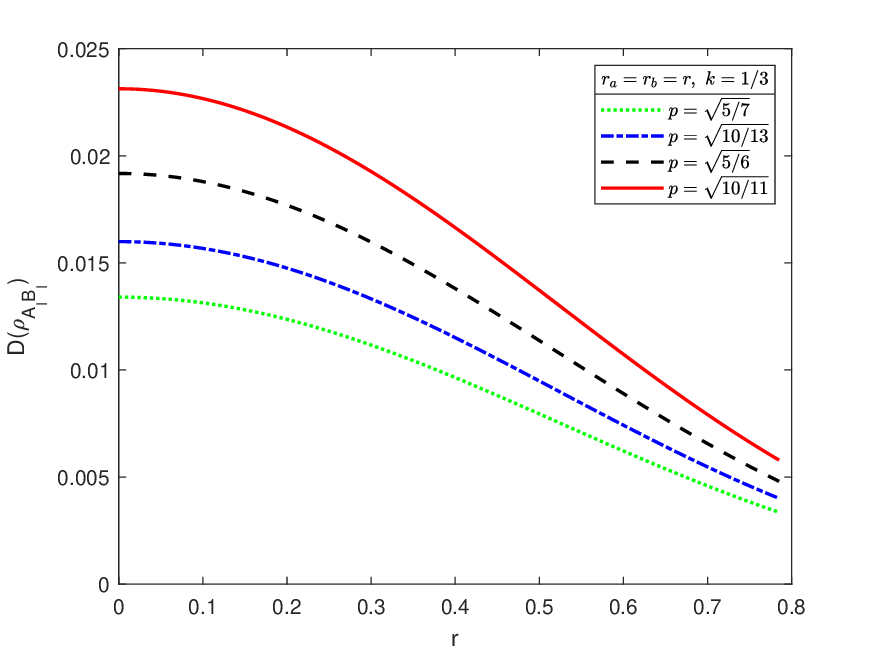}
		\caption{  }
		\label{fig:subfig1}
	\end{subfigure}
  \begin{subfigure}[h]{0.3\textwidth}
		\includegraphics[width=\textwidth]{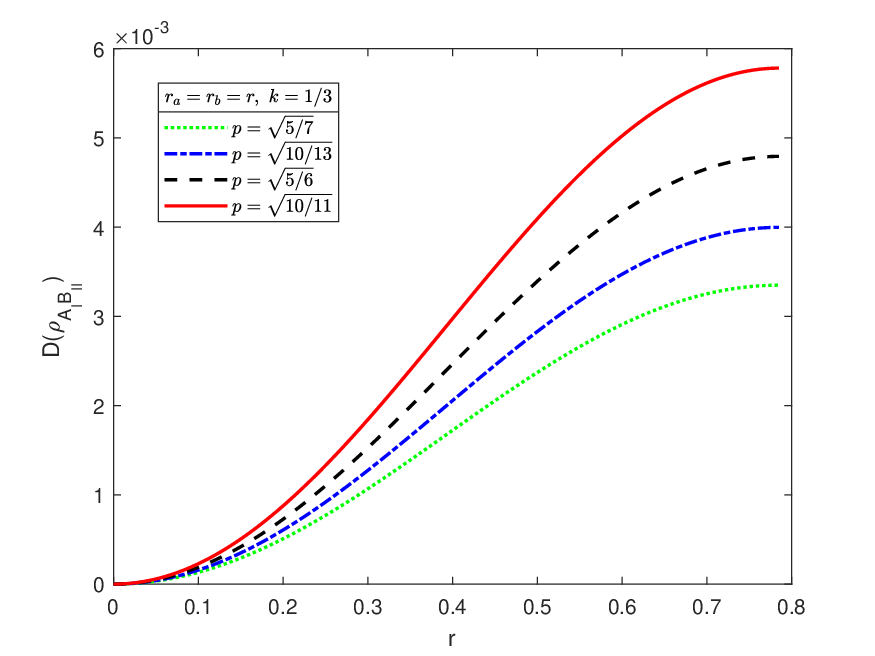}
		\caption{ }
		\label{fig:subfig2}
	\end{subfigure}
  \begin{subfigure}[h]{0.3\textwidth}
		\includegraphics[width=\textwidth]{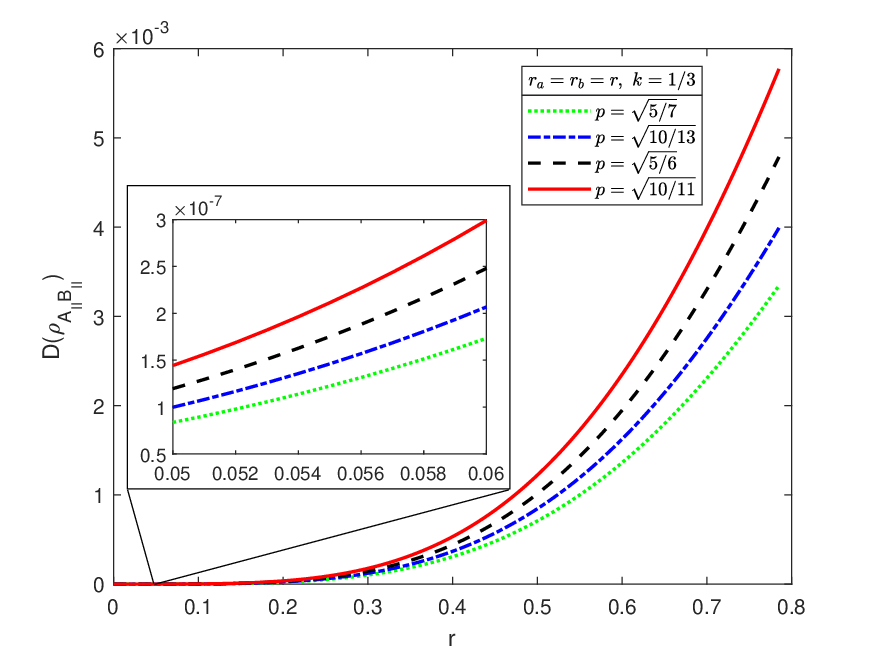}
		\caption{  }
		\label{fig:subfig2}
	\end{subfigure}

\begin{subfigure}[h]{0.3\textwidth}
		\includegraphics[width=\textwidth]{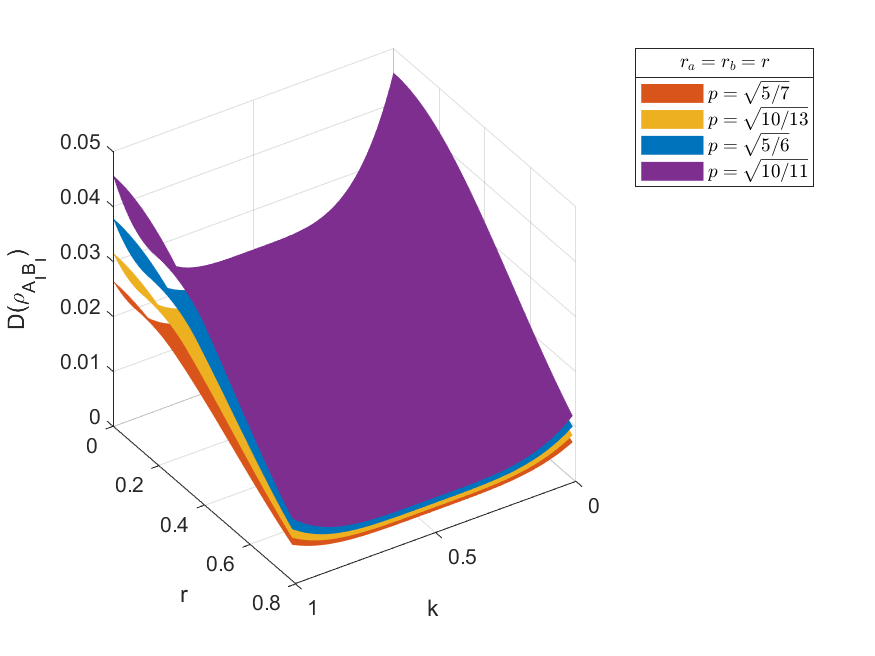}
		\caption{ }
		\label{fig:subfig1}
	\end{subfigure}
  \begin{subfigure}[h]{0.3\textwidth}
		\includegraphics[width=\textwidth]{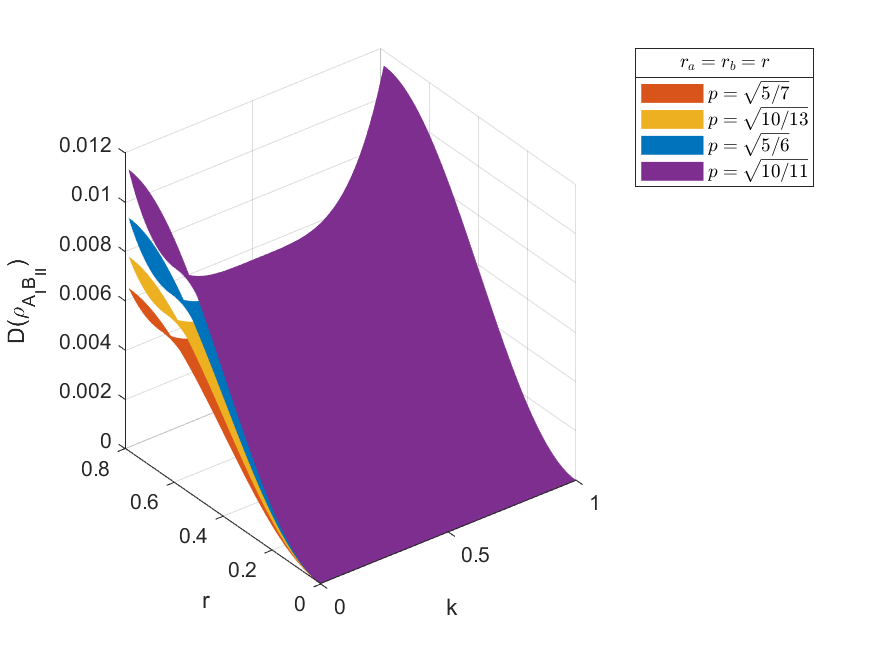}
		\caption{ }
		\label{fig:subfig2}
	\end{subfigure}
  \begin{subfigure}[h]{0.3\textwidth}
		\includegraphics[width=\textwidth]{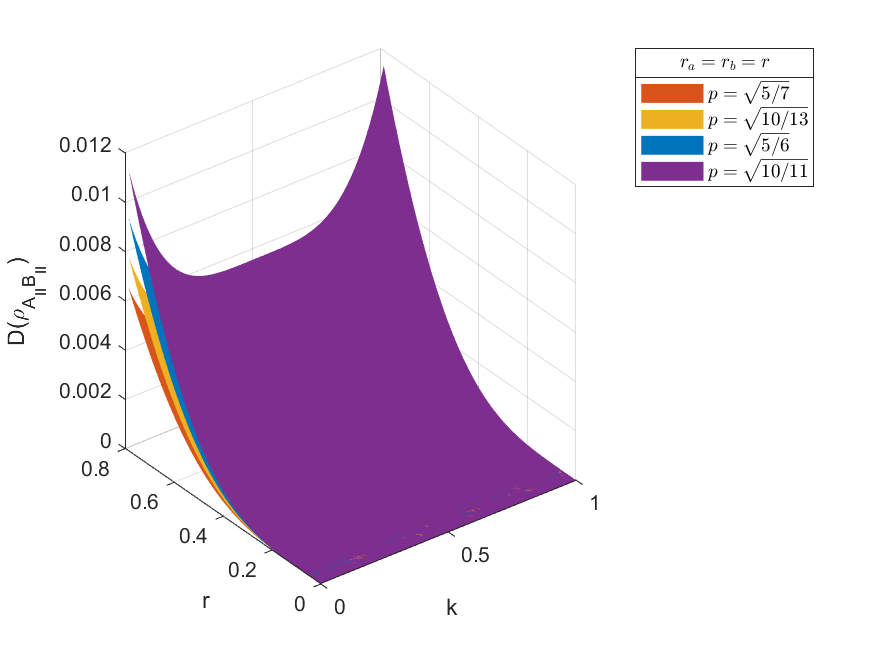}
		\caption{  }
		\label{fig:subfig2}
	\end{subfigure}	
	\label{Fig.6}\caption{Plot quantum discords $D(\rho_{A_{I}B_I})$, $D(\rho_{A_{I}B_{II}} )$ and $D(\rho_{A_{II}B_{II}} )$ for the bit flip channel when $r_a=r_b=r$. The upper subfigures are the cases that discord is a function of acceleration parameter $r$  for $k=\frac{1}{3}$. The lower subfigures are the cases that discord is a function of both acceleration parameter $r$ and decay probability parameter $k$. }
\end{figure}
However, both the noisy environment
and the acceleration of subsystems influence the geometric measure of quantum discords,
but the geometric measure of quantum discords sudden death never occurs.

\section{Conclusions}
In this work, based on the Werner state, shared by Alice and Bob, the decoherence of discord in non-inertial frames is investigated under noisy channels such as phase flip, bit flip and phase damping channels. We have derived analytical geometric measure of quantum discords of Dirac fields in noninertial frames and discussed their behaviors in terms of the acceleration $r$. Serval trade-off relations of quantum discords about reduced density matrices are proposed. It is shown that both the noisy environment and the acceleration of subsystems influence the quantum discord, but, for any noisy environment and for
any acceleration, the discord sudden death never occurs.

\bigskip
\noindent{\bf Acknowledgments}
This work is supported by the National Natural
Science Foundation of China (NSFC) under Grant No. 12204137; the China Scholarship Council (CSC); and
the Natural Science Foundation of Hainan Province under Grant No. 125RC744.

\section{Appendix }
It is worth noting that the linear map given by
(\ref{R1}) and (\ref{R2}), exactly, induces a linear map on the operator space from $\mathcal{C}$  to $\mathcal{C}\otimes \mathcal{C}$ (here, we call it $\Lambda$ ).
Specifically, the map $\Lambda$ acts on the unit matrices $\{\ket{0}\bra{0}, \ket{0}\bra{1}, \ket{1}\bra{0}, \ket{1}\bra{1}\}$,  it has representations as follows,
\begin{equation}\begin{aligned}\label{R4}
\Lambda(\ket{0}\bra{0})&= \cos^2r \ket{0}\bra{0}_I\otimes \ket{0}\bra{0}_{II}+\cos r\sin r \ket{0}\bra{1}_I\otimes \ket{0}\bra{1}_{II}\\
&+\sin r\cos r \ket{1}\bra{0}_I\otimes \ket{1}\bra{0}_{II}+\sin^2 r \ket{1}\bra{1}_I\otimes \ket{1}\bra{1}_{II},\\
\Lambda(\ket{0}\bra{1})&= \cos r \ket{0}\bra{1}_I\otimes \ket{0}\bra{0}_{II}+\sin r \ket{1}\bra{1}_I\otimes \ket{1}\bra{0}_{II},\\
\Lambda(\ket{1}\bra{0})&= \cos r \ket{1}\bra{0}_I\otimes \ket{0}\bra{0}_{II}+\sin r \ket{1}\bra{1}_I\otimes \ket{0}\bra{1}_{II},\\
\Lambda(\ket{1}\bra{1})&= \ket{1}\bra{1}_I\otimes \ket{0}\bra{0}_{II}.
\end{aligned}
\end{equation}
However, the Pauli matrices $\{\sigma_j\}$, where $j=0, 1, 2, 3$, and $\sigma_0=I$, as another famous basis of operator space $\mathcal{C}$, they have different representations under the map $\Lambda$, that is,
\begin{equation}\label{R5}
\begin{aligned}
\Lambda (I)= &\cos^2r \ket{0}\bra{0}_I\otimes \ket{0}\bra{0}_{II}+\cos r\sin r \ket{0}\bra{1}_I\otimes \ket{0}\bra{1}_{II}+\sin r\cos r \ket{1}\bra{0}_I\otimes \ket{1}\bra{0}_{II}\\
&+\sin^2 r \ket{1}\bra{1}_I\otimes \ket{1}\bra{1}_{II}+\ket{1}\bra{1}_I\otimes \ket{0}\bra{0}_{II},\\
\Lambda (\sigma_1)= &\cos r (\ket{0}\bra{1}_{I}+\ket{1}\bra{0}_{I})\otimes \ket{0}\bra{0}_{II}+\sin r \ket{1}\bra{1}_{I}\otimes (\ket{0}\bra{1}_{II}+\ket{1}\bra{0}_{II}),\\
\Lambda (\sigma_2)= & \texttt{i}\cos r (\ket{1}\bra{0}_{I}-\ket{0}\bra{1}_{I})\otimes \ket{0}\bra{0}_{II}+\texttt{i} \sin r \ket{1}\bra{1}_{I}\otimes (\ket{0}\bra{1}_{II}-\ket{1}\bra{0}_{II}),\\
\Lambda (\sigma_3)= &\cos^2r \ket{0}\bra{0}_I\otimes \ket{0}\bra{0}_{II}+\cos r\sin r \ket{0}\bra{1}_I\otimes \ket{0}\bra{1}_{II}+\sin r\cos r \ket{1}\bra{0}_I\otimes \ket{1}\bra{0}_{II}\\
&+\sin^2 r \ket{1}\bra{1}_I\otimes \ket{1}\bra{1}_{II}-\ket{1}\bra{1}_I\otimes \ket{0}\bra{0}_{II}.
\end{aligned}
\end{equation}

In fact, Eqs. (\ref{R4}) and (\ref{R5}) offer an effective method to understand the Unruh effect from the operator viewpoint, since any operator on $\mathcal{C}$ can be rewritten as a linear combination of Pauli matrices (i.e., the Bloch representation), it is helpful to discuss the quantum correlation of any quantum state under the Hawking radiation.

Since the Pauli matrices  $\{\sigma_i\}$ ($i=0, 1, 2, 3$) can be decomposed according to the operator basis $\{\ket{0}\bra{0}, \ket{0}\bra{1}, \ket{1}\bra{0}, \ket{1}\bra{1}\}$ as follows,
\begin{equation}
\begin{aligned}
  I &= \ket{0}\bra{0}+\ket{1}\bra{1} \\
  \sigma_1 &= \ket{0}\bra{1}+\ket{1}\bra{0} \\
  \sigma_2 &= -i\ket{0}\bra{1}+i\ket{1}\bra{0} \\
  \sigma_3 &= \ket{0}\bra{0}-\ket{1}\bra{1}.
\end{aligned}
\end{equation}
Employing map (\ref{R4}), the Pauli matrices are deformed into two modes, that is,
\begin{equation}\label{R5}
\begin{aligned}
\Lambda (I)= &\cos^2r \ket{0}\bra{0}_I\otimes \ket{0}\bra{0}_{II}+\cos r\sin r \ket{0}\bra{1}_I\otimes \ket{0}\bra{1}_{II}+\sin r\cos r \ket{1}\bra{0}_I\otimes \ket{1}\bra{0}_{II}\\
&+\sin^2 r \ket{1}\bra{1}_I\otimes \ket{1}\bra{1}_{II}+\ket{1}\bra{1}_I\otimes \ket{0}\bra{0}_{II},\\
\Lambda (\sigma_1)= &\cos r (\ket{0}\bra{1}_{I}+\ket{1}\bra{0}_{I})\otimes \ket{0}\bra{0}_{II}+\sin r \ket{1}\bra{1}_{I}\otimes (\ket{0}\bra{1}_{II}+\ket{1}\bra{0}_{II}),\\
\Lambda (\sigma_2)= & \texttt{i}\cos r (\ket{1}\bra{0}_{I}-\ket{0}\bra{1}_{I})\otimes \ket{0}\bra{0}_{II}+\texttt{i} \sin r \ket{1}\bra{1}_{I}\otimes (\ket{0}\bra{1}_{II}-\ket{1}\bra{0}_{II}),\\
\Lambda (\sigma_3)= &\cos^2r \ket{0}\bra{0}_I\otimes \ket{0}\bra{0}_{II}+\cos r\sin r \ket{0}\bra{1}_I\otimes \ket{0}\bra{1}_{II}+\sin r\cos r \ket{1}\bra{0}_I\otimes \ket{1}\bra{0}_{II}\\
&+\sin^2 r \ket{1}\bra{1}_I\otimes \ket{1}\bra{1}_{II}-\ket{1}\bra{1}_I\otimes \ket{0}\bra{0}_{II}.
\end{aligned}
\end{equation}

Now, in order to obtain the subsystem, we trace out $II$ mode, and we have the operators on $I$ mode,
\begin{equation}\label{R6}
\begin{aligned}
(I)_I&=\cos^2 r \ket{0}\bra{0}+\sin^2 r\ket{1}\bra{1}+\ket{1}\bra{1}=I-\sin^2 r\sigma_3,\\
(\sigma_1)_I&=\cos r\ket{0}\bra{1}+\cos r\ket{1}\bra{0}=\cos r \sigma_1,\\
(\sigma_2)_I&=i \cos r\ket{1}\bra{0}-i\cos r\ket{0}\bra{1}=\cos r \sigma_2,\\
(\sigma_3)_I&=\cos^2 r\ket{0}\bra{0}+\sin^2 r\ket{1}\bra{1}-\ket{1}\bra{1}=\cos^2 r\sigma_3.\\
\end{aligned}
\end{equation}

Similarly, tracing out $I$ mode, we have the operators on  $II$ mode
\begin{equation}\label{R7}
\begin{aligned}
(I)_{II}&=\cos^2 r \ket{0}\bra{0}+\sin^2 r\ket{1}\bra{1}+\ket{0}\bra{0}=I+\cos^2 r\sigma_3,\\
(\sigma_1)_{II}&=\sin r\ket{0}\bra{1}+\sin r\ket{1}\bra{0}=\sin r \sigma_1,\\
(\sigma_2)_{II}&=i \sin r\ket{0}\bra{1}-i\sin r\ket{1}\bra{0}=-\sin r\sigma_2,\\
(\sigma_3)_{II}&=\cos^2 r\ket{0}\bra{0}+\sin^2 r\ket{1}\bra{1}-\ket{0}\bra{0}=-\sin^2 r\sigma_3.\\
\end{aligned}
\end{equation}

For any bipartite quantum state, it has Bloch representation like this,
\begin{equation}
\rho=\frac{1}{4}(I_A\otimes I_B+\sum_i x_i \sigma_i\otimes I_B+\sum_j y_jI_A\otimes \sigma_j+\sum_{i,j=1}^3t_{ij}\sigma_i\otimes \sigma_j),
\end{equation}
 Now we consider both Alice and Bob hover near the event horizon of a Schwarzschild black hole, under the transforms (\ref {R5}), the initial bipartite quantum state should be transformed a four-partite quantum state, then using Eqs. (\ref{R6}) and (\ref{R7}), it is easy to obtain the partial trace quantum states,
\begin{equation}
\begin{aligned}
\rho_{A_IB_I}=&\frac{1}{4}(I_A\otimes I_B-\sin^2 r_a  \sigma_3\otimes I_B-\sin^2 r_b I_A\otimes \sigma_3+\frac{2p-1}{3}\cos r_a\cos r_b\sigma_1\otimes \sigma_1\\
&+\frac{2p-1}{3}\cos r_a\cos r_b\sigma_2\otimes \sigma_2
+(\sin^2 r_a\sin^2 r_b+\frac{2p-1}{3}\cos^2 r_a\cos^2 r_b)\sigma_3\otimes \sigma_3),
\end{aligned}
\end{equation}

\begin{equation}
\begin{aligned}
\rho_{A_IB_{II}}=&\frac{1}{4}(I_A\otimes I_B-\sin^2 r_a  \sigma_3\otimes I_B+\cos^2 r_b I_A\otimes \sigma_3+\frac{2p-1}{3}\cos r_a \sin r_b\sigma_1\otimes \sigma_1\\&-\frac{2p-1}{3}\cos r_a\sin r_b\sigma_2\otimes \sigma_2
-(\sin^2 r_a\cos^2 r_b+\frac{2p-1}{3}\cos^2 r_a\sin^2 r_b)\sigma_3\otimes \sigma_3),
\end{aligned}
\end{equation}

\begin{equation}
\begin{aligned}
\rho_{A_{II}B_{I}}=&\frac{1}{4}(I_A\otimes I_B+\cos^2 r_a  \sigma_3\otimes I_B-\sin^2 r_b I_A\otimes \sigma_3+\frac{2p-1}{3}\sin r_a \cos r_b\sigma_1\otimes \sigma_1\\&-\frac{2p-1}{3}\sin r_a\cos r_b\sigma_2\otimes \sigma_2
-(\cos^2 r_a\sin^2 r_b+\frac{2p-1}{3}\sin^2 r_a\cos^2 r_b)\sigma_3\otimes \sigma_3),
\end{aligned}
\end{equation}

\begin{equation}
\begin{aligned}
\rho_{A_{II}B_{II}}=&\frac{1}{4}(I_A\otimes I_B+\cos^2 r_a  \sigma_3\otimes I_B+\cos^2 r_b I_A\otimes \sigma_3+\frac{2p-1}{3}\sin r_a\sin r_b\sigma_1\otimes \sigma_1\\
&+\frac{2p-1}{3}\sin r_a\sin r_b\sigma_2\otimes \sigma_2
+(\cos^2 r_a\cos^2 r_b+\frac{2p-1}{3}\sin^2 r_a\sin^2 r_b)\sigma_3\otimes \sigma_3),
\end{aligned}
\end{equation}

%

%
%


\begin{thebibliography}{00}

\bibitem{app1}S. Luo, Using measurement-induced disturbance to characterize correlations as classical or quantum, Phys. Rev. A {\bf77}, 022301 (2008).
\bibitem{app2}N. Li and S. Luo, Classical states versus separable states, Phys. Rev. A {\bf78}, 024303 (2008).
 \bibitem{app3}C. Radhakrishnan, M. Lauriere, and T. Byrnes, A multipartite generalization of quantum discord, Phys. Rev. Lett. {\bf124}, 110401 (2020).

 \bibitem{app4}Y. Inui and Y. Yamamoto, Entanglement and quantum discord in optically coupled coherent Ising machines, Phys. Rev. A {\bf102}, 062419 (2020).
 \bibitem{hu2020}J. Zhou, X. Hu, and N. Jing, Quantum discord of certain two-qubit states, Int. J. Theo. Phys. {\bf59}, 415(2020).

\bibitem{def1}C.C. Rulli and M.S. Sarandy, Global quantum discord in multipartite systems, Phys. Rev. A {\bf84}, 042109 (2011).
\bibitem{def2}S. Luo and S. Fu, Geometric measure of quantum discord, Phys. Rev. A {\bf82}, 034302 (2010).

 \bibitem{olli}H. Ollivier and W. H. Zurek, Quantum discord: a measure of the quantumness of correlations, Phys. Rev. Lett. \textbf{88}, 017901 (2001).

\bibitem{entangle1}X. Liu, C. Zeng, and J. Wang, Generation of quantum entanglement in superposed diamond spacetime, arXiv:2501.00246.

\bibitem{entangle2}Q. Liu, S. Wu, C. Wen, and J. Wang, Quantum properties of fermionic fields in multi-event horizon spacetime, Sci. China-Phys. Mech. Astron. {\bf66}, 120413 (2023).
\bibitem{entangle3}J. Wang, C. Wen, S. Chen, and J. Jing, Generation of genuine tripartite entanglement for continuous variables in de Sitter space, Phys. Lett. B {\bf800}, 135109 (2020).
 \bibitem{entangle4}G. Mi, X. Huang, S. Fei, and T. Zhang, Impact of the Hawking effect on the fully entangled fraction of three-qubit states in Schwarzschild spacetime, Ann. Phys. (Berlin), 2400308(2024).

\bibitem{coherence1}W. Li, J. Lu, and S. Wu, Multiqubit coherence of mixed states near event horizon, arXiv:2505.07476.
 \bibitem{coherence2}S. Wu, H. Zeng, and H. Cao, Quantum coherence and distribution of N-partite bosonic fields in noninertial frame, arXiv.2201.00986.
 \bibitem{coherence3}T. Liao, J. Yang, T. Zhang, and X. Huang, Quantum coherence of mixed states under noisy channels in noninertial frames, Results in Phys. {\bf70}, 108169 (2025).
 \bibitem{mi2025}G. Mi, X. Huang, S. Fei, and T. Zhang, Genuine four-partite Bell nonlocality in the curved spacetime, Eur. Phys. J. C {\bf85}, 354(2025).
\bibitem{zhang2023}T. Zhang, X. Wang, and S. Fei, Hawking effect can generate physically inaccessible genuine tripartite nonlocality, Eur. Phys. J. C {\bf83}, 607(2023).
\bibitem{steering1}S. Wu, J. Li, Y. Wang, S. Shang, and J. Lu, Fermionic steering in multi-event horizon spacetime, Eur. Phys. J. C {\bf85}, 54(2025).
\bibitem{steering2}S. Wu, H. Wu, Y. Wang, and J. Wang, Gaussian tripartite steering in Schwarzschild black hole, Phys. Lett. B {\bf865}, 139493 (2025).
\bibitem{steering3}T. Liu, J. Wang, J. Jing, and H. Fan, The influence of Unruh effect on quantum steering for accelerated two-level detectors with different measurements, Ann. Phys. {\bf390}, 334(2018).

\bibitem{liu2018}T. Liu, J. Jing, and J. Wang, Satellite-based quantum steering under the influence of spacetime curvature of the Earth, Adv. Quantum Technol. {\bf1}, 1800072(2018).
\bibitem{liu2019}T. Liu, S. Cao, S. Wu, and H. Zeng, The influence of the Earth¡¯s curved
spacetime on Gaussian quantum coherence, Laser Phys. Lett. {\bf16}, 095201(2019).
\bibitem{wang2011}J. Wang and J. Jing, Multipartite entanglement of fermionic systems in noninertial frames, Phys. Rev. A {\bf83}, 022314 (2011).
\bibitem{torres2019}A.J. Torres-Arenasa, Q. Dong, G.H. Sun, W.C. Qiang, and S.H. Dong, Entanglement measures of W-state in noninertial frames,  Phys. Lett. B {\bf789}, 93(2019).
\bibitem{wu2020}S. Wu, Z. Li, and H.  Zeng, Quantum steering between two accelerated
parties, Laser Phys. Lett. {\bf17}, 035202(2020).

\bibitem{unrh1}Z. Huang and H. Situ, Quantum coherence behaviors of fermionic systems in non-inertial frame, Quantum Inf. Process. {\bf 17}, 95(2018).
\bibitem{unrh2}Z.Y. Ding, C.C. Liu, W.Y. Sun, J. He, and L. Ye, Quantum coherence of fermionic systems in noninertial frames beyond the single-mode approximation, Quantum Inf. Process. {\bf17}, 279 (2018).

\bibitem{unrh3}P.M. Alsing, I. Fuentes-Schuller, R.B. Mann, and T.E. Tessier, Entanglement of Dirac fields in noninertial frames, Phys. Rev. A {\bf74}, 032326(2006).
 \bibitem{vedral}L. Henderson and V. Vedral, Classical, quantum and total correlations, J. Phys. A {\bf34}, 6899 (2001).
 \bibitem{luo2008}S. Luo, Quantum discord for two-qubit systems, Phys. Rev. A {\bf77}, 042303 (2008).
\bibitem{zhu2021}C. Zhu, B. Hu, B. Li, Z. Wang, and S. Fei, Geometric discord for multiqubit systems, arXiv:2104.12344.
\bibitem{libo2021}B. Li, C. Zhu, X. Liang, B. Ye, and S. Fei, Quantum discord for multiqubit systems, Phys. Rev. A {\bf104}, 012428 (2021).
\bibitem{hunt2019}M.A. Hunt, I.V. Lerner, I.V. Yurkevich, and Y. Gefen, How to observe and quantify quantum-discorded states, Phys. Rev. A {\bf100}, 022321 (2019).
\bibitem{aaron2019}A. Szasz, A measure of quantum correlations that lies approximately between entanglement and discord, Phys. Rev. A {\bf99}, 062313 (2019).
\bibitem{hu2025}J. Zhou, X. Hu, H. Zhang, and N. Jing, Analytic formulas for quantum discord of special families of n-qubit states, Quant. Inf. Process. 24, (2025).

\bibitem{dis2010}B. Daki$\acute{c}$, V. Vedral, and $\check{C}$ Brukner, Necessary and sufficient condition for nonzero quantum discord, Phys. Rev. Lett. {\bf105}, 190502 (2010).
\bibitem{werner}R.F. Werner, Quantum states with Einstein-Podolsky-Rosen correlations admitting a hidden-variable model, Phys. Rev. A {\bf40}, 4277 (1989).
\bibitem{decay}M.A. Nielsen and I.L. Chuang, Quantum computation and quantum information (Cambridge: Cambridge University Press) 2010.




\end{thebibliography}
\end{document}